# Chiral π Domain Walls Composed of Twin Half-Integer Surface Disclinations in Ferroelectric Nematic Liquid Crystals

Shengzhu Yi[1,4], Zening Hong[1], Zhongjie Ma[1], Chao Zhou[1], Miao Jiang[1], Xiang Huang[2], Mingjun Huang[2,3], Satoshi Aya[2,3], Rui Zhang[4], Qi-Huo Wei[1,5,*]

[1] Department of Mechanical and Energy Engineering, Southern University of Science and Technology, Shenzhen, 518055, China.

[2] South China Advanced Institute for Soft Matter Science and Technology (AISMST), School of Emergent Soft Matter, South China University of Technology, Guangzhou 510640, China.

[3] Guangdong Provincial Key Laboratory of Functional and Intelligent Hybrid Materials and Devices, South China University of Technology, Guangzhou 510640, China.

[4] Department of physics, The Hongkong University of Science and Technology, Clear Water Bay, Hong Kong, China.

[5] Center for Complex Flows and Soft Matter Research, Southern University of Science and Technology, Shenzhen 518055, China.

*Corresponding author. Email: weiqh@sustech.edu.cn

**Abstract**

Ferroelectric nematic liquid crystals are polar fluids characterized by microscopic orientational ordering and macroscopic spontaneous polarizations. Within these fluids, walls that separate domains of different polarizations are ubiquitous. We demonstrate that the π walls in films of polar fluids consist of twin half-integer surface disclinations spaced horizontally, enclosing a subdomain where the polarization exhibits left- or right-handed π twists across the film. The degenerate geometric configurations of these twin disclinations give rise to kinks and antikinks, effectively partitioning subdomains of opposite chirality like Ising chains. The hierarchical topological structures dictate that field-driven polar switching entails a two-step annihilation process of the disclinations. These findings serve as a cornerstone for comprehending other walls in ferroelectric and ferromagnetic materials, thereby laying the base for domain engineering crucial for advancing their nonlinear and optoelectronic applications.

**Introduction**

Over a century ago, Debye[1] and Born[2] envisioned an intriguing state of matter, wherein anisotropic molecules align their electrical dipoles in a common direction, manifesting macroscopically spontaneous polarizations **P** while maintaining high fluidity. This polar fluid, referred to as a ferroelectric nematic ($N_F$) liquid crystal, has recently been materialized and demonstrated[3–5]. In contrast to crystalline ferroelectric materials having broken translational symmetry, $N_F$ liquid crystals possess continuous translation symmetry. $N_F$ liquid crystals exhibit remarkable properties, including low driving field[6–8], high optical nonlinear response[9,10], polar topological structures[11,12], and a chiral ground state[13,14], which are not only of scientific interest but also of practical importance.

π walls are interfaces where adjacent polarizations are anti-parallel with each other[15,16]. The Ginzburg–Landau theory ascertains that the ferroelectric π wall is of the Ising type with a hyperbolic tangent (tanh) profile of the polarization[17]. This contrasts with domain walls in ferromagnetic materials, where magnetization reversal can only be achieved via spin rotation, as the spins are quantized. Thus, ferromagnetic domain walls are the Bloch (or Néel) type when the rotation plane of the spins is parallel (or perpendicular) to the wall[15]. The common view—that the ferroelectric π walls are of the Ising type—has rarely been challenged, except that recent studies suggest the possibility of mixed Ising-Néel/Bloch walls[18–20]. Despite significant advances in understanding π walls in solid ferroic materials, the internal structures of their fluid counterparts ($N_F$) remain poorly understood. For example, the π walls feature two lines and step-like variations, a phenomenon observed but unexplained in prior studies[21–23]. Recently, a pioneering study suggests that π walls are characterized by a continuous rotation of polarization without disclinations[24]. The suggested structures of domain walls, however, contradicts the picture that they are bend textures accompanied by two overlapping surface disclinations[25].

Here, we report studies on the structures of π walls in slabs of $N_F$ liquid crystals confined between two treated substrates to induce a planar uniaxial alignment. The $N_F$ liquid crystals produce stripe domains with alternating polarizations separated by π walls, and the mean domain size exhibits a linear dependence on the cell thickness, violating Kittel's law. The π walls consist of twin half-integer surface disclinations that are separated horizontally, enclosing subdomains where the polarization exhibits twists across the cell. Two degenerate configurations of the twin surface disclinations lead to excitations of kinks and antikinks along π walls, parting subdomains with opposite chirality. This hierarchical topology of π walls dictates that the field-driven polar switching undergoes two annihilation steps of the surface disclinations.

**Structures and Scaling Laws for Domains and π Walls**

We spin-coated thin films of polyimide on glass substrates and mechanically rubbed them along a direction, denoted as **R.** We assembled the treated substrates, in a parallel manner, into liquid crystal cells. The cell thickness, denoted as $h$, was controlled within the range of 1 μm to 10 μm using silica spacers. The detailed fabrication processes for the cells are provided in the Supplementary Information (Extended Data Fig. 1). The cells were filled with ferroelectric materials at their isotropic (I) phase and subsequently cooled to the $N_F$ phase at a cooling rate of 5 K/min. We used two $N_F$ materials: RM734 and DIO, that were first reported by Mandle et al.[4] and Nishikawa et al.[3] respectively.

The phase sequences of DIO and RM734 are I − 174 °C − Nematic (N) − 66°C − $N_F$ − Crytal, and I − 187 °C − N − 133°C − $N_F$ − Crytal, respectively. In the N phase, the cell shows a monodomain texture (Supplementary Video 1); the rod-like molecules orient, on average, in a common direction, called the director (**n**). When the liquid crystal enters the $N_F$ phase, it generates stripes aligned parallel to the rubbing direction (Fig. 1a-d and Extended Data Fig. 2). As the electrical dipoles are along the long axes of these molecules, the spontaneous polarization **P** is parallel to **n**.

Figures 1a-d show polarized optical microscopy (POM) images of the ferroelectric domains. When crossed polarizers are parallel or perpendicular to **R**, the disappearance of the transmitted light suggests that the molecular orientations are parallel or antiparallel to **R** (Fig. 1b). The molecular orientations were verified by measuring the slow optical axis using the PolScope technique (Extended Data Fig. 3). We discerned the polarization directions by applying an in-plane electric field **E** (Extended Data Fig. 4). The domains that remained unaffected had their polarizations parallel to **E**, while their adjacent domains, which underwent molecular reorientation, had their polarizations antiparallel to **E**. Thus, the stripes have alternating polarizations, and the domain boundaries are π walls.

A distinctive characteristic of π walls is their configuration of two parallel lines with a uniform but finite separation. One line appears in bright, while the other appears dark (Fig. 1a and Extended Data Fig. 5). When a slight pressure (~1 kPa) was applied on the top surface of the cell, the distance between these paired lines increased/decreased (Fig. 1e and Supplementary Video 2); however, after the pressure was released, the lines returned to their original state. Based on this observation, we speculate that the π wall consists of twin lines positioned at two confining surfaces.

Another notable aspect of π walls is the areas bounded by the twin lines, referred to as subdomains (Fig. 1a). Particularly, when the sample is rotated under the crossed polarizers, the intensity of the subdomains remains largely unchanged, indicating a twist of the polarization across the cell (Fig. 1a-b). The chirality of the subdomain was further validated by rotating the polarizer with respect to the analyzer (Fig. 1c, d and Extended Data Fig. 6). The twist subdomain becomes particularly visible when a π wall splits into two walls (see dashed lines in Fig. 1b-d), where one subdomain grows into a larger twist domain and maintains the same POM texture (so the chirality). The twist can be either right-handed (RH) or left-handed (LH), thus, the π walls appear differently under uncrossed polarizers (Fig. 1a-d).

To deduce the polarization orientations across the π walls, we used a second-harmonic generation (SHG) microscope. When circularly polarized light is applied perpendicularly to the cell, the π walls exhibit two dark lines. This result indicates either that the polarization (**P**) is perpendicular to the cell or that magnitude of the polarization is suppressed[11,26]. Since a wall with a diminishing |**P**| and no director rotations is scarcely visible under an optical microscopy[27], the absence of second-harmonic emissions is primarily attributed to the out-of-plane tilt of the polarizations. This tilt is consistent with the result of a previous experiment[24].

Both the domain width $w$ and the twin line separation $d_0$ are influenced by the cell thickness. Experimental data demonstrate that these parameters grow linearly with increasing cell thickness (Fig. 1g-h). Notably, linear fitting with $w = kh$ yields the prefactors $k = 2.95$ for RM734 and $k = 2.05$ for DIO. This linear relationship differs from the square root dependence, $w \sim \sqrt{h}$, known as Kittel's law, which is universal for solid-state ferroic materials[15,28].

The deviation from Kittel's law can be ascribed to the formation of domain walls in $N_F$ materials during the ferroelectric-ferroelastic phase transition[26]. The transition is driven by an instability toward splay deformations via flexoelectric coupling: $-\gamma \mathbf{n}(\boldsymbol{\nabla} \cdot \mathbf{n}) \cdot \mathbf{P}$, where $\gamma$ represents the splay flexoelectric coefficient. The ensuing splay deformation is periodic and alternating in directions, as a homogeneous splay is incapable of occupying the entire space. The deformation period is linearly related to the cell thickness and increases as the temperature decreases[29]. When the cell enters the $N_F$ phase, the period of the modulated polarization is preserved, as observed in the experiment (Supplementary Video 1)[30].

The alignment method affects the characteristics of ferroelectric domains. In cells where the rubbing directions at the two surfaces are parallel, the average width of the domains with **P** parallel to **R** is ~25% greater than that of the domains with **P** antiparallel to **R** (Fig.1e and Extended Data Fig. 2). This difference arises because the rubbed surface induces polar anchoring, favoring the alignment of **P** with **R**[24,31]. Conversely, in cells where the rubbing directions on the two surfaces are antiparallel, the domains display nonuniform sizes and twists of polarization across the cell[32] (Extended Data Fig. 7). In cells made with photoalignment method, the domains appeared less uniform in size and revealed larger widths than those in the rubbed cells (Extended Data Fig. 2 and Extended Data Fig. 8).

**Topological Structures of the π Walls**

Based on these observations, we propose that the π walls consist of two surface lines adopting a Bloch-type configuration (Fig. 2a). When the two lines are horizontally separated, the subdomain manifests antiparallel polarizations at two confining surfaces, leading to a π-twist of the polarization along the cell normal. We calculated the polarization fields using a modified Ericksen-Leslie model[33,34]. The details of the simulations are described in the Methods.

We calculated the total free energies for different line separations, namely: $d = X_t - X_b$, where $X_t$ and $X_b$ represent the locations of the lines at the top and bottom surfaces, respectively. For a fixed cell thickness, the total energy displays two minima at $d = \pm d_0$, representing two degenerate configurations (Fig. 2b). The calculated optimal separation $d_0$ increases linearly with the cell thickness, which is in good agreement with the experimental findings. For the optimal line separation, we calculated the polarization field and reproduced the optical textures observed in the experiments (Extended Data Fig. 9).

The free energy reaches a maximum at $d = 0$, indicating that overlaps of the two domain lines is energetically unfavorable (Fig. 2b). This is primarily attributed to the sharp increase in the bulk energy. The underlying physics are rooted in the conflict between the polarization directions at the twin lines, leading to a significant decrease in the polar order parameter (insert in Fig. 2c). The emergence of the double minima arises from the interplay of elastic energies between the associated splay and twist distortions (Fig. 2c). They display opposite dependencies on the line separation, while the bend and bulk energies exhibit minimal variations with the line separation.

The calculated polarization field shows that the π walls encompass the Ising, Bloch, and Néel characteristics. Specifically, near the confining surface, the polarization reverses by decreasing its magnitude and rotating out-of-plane, which are the signatures of the Ising and Bloch walls[35], respectively (Fig. 2d). However, in the middle plane of the cell, the polarization reverses via a smooth in-plane rotation, which is a hallmark of the Néel wall (Fig. 2e).

These two surface lines are two disclinations. To determine their topological features, we consider a half-circle centered at the disclinations (Fig. 2f) and map the associated polarizations onto the ground-state manifold (Fig. 2g). The manifold for the $N_F$ liquid crystal is a spherical surface ($\mathbb{S}^2$), where each point represents a degenerate polarization orientation[36,37]. The mapping forms a half-circle on either the upper or lower hemisphere of the order parameter space. Following Kleman's tradition[38], the topological charge of the surface disclination is defined as: $Q_w = \Delta\varphi/2\pi$, where $\Delta\varphi$ is the angle difference between the start and end of the half-circle. Thus, the topological strengths of the surface disclinations become: $Q_w = \pm\pi/2\pi = \pm1/2$, the twin disclinations of an $\pi$ wall have identical topological charges, and the disclinations of two adjacent $\pi$ walls have opposite charges.

**Kinks along the π Walls**

Figure 3a present polarized optical microscopic images of the step-like variations along π walls. We denoted the normalized horizontal distance between twin disclinations as the order parameter: $\phi(y) = [X_t(y) - X_b(y)]/d_0$ and found that it can be well fitted by using the tangent kink function: $\phi(y) = \tanh(y/\xi)$, where $\xi$ is the fitting parameter (Fig. 3b). Excitations of such kinks can be attributed to the degenerate configuration of the twin disclinations. The double well potential, exhibited by the free energy of the π wall, is similar to the $\phi^4$ model[39]. The kinks are categorized as $\mathbb{Z}_2$ kinks due to the invariance of the free energy after transformation $F(d) \rightarrow F(-d)$ (Fig. 2b).

Kinks are topological excitations in one-dimensional systems, with topological charges defined as: $Q_k = [\phi(y^+) - \phi(y^-)]/2 = \pm1$, where $\phi(y^\pm)$ represents the order parameter at two sides of the kink. Antikinks have a charge $Q_k = -1$. The total topological charge is a conservation quantity[39]. The states of unclosed π walls can be represented by $[\phi(-\infty), \phi(\infty)]$, then we have four different states: $[1, 1]$ and $[-1, -1]$ for $\sum Q_k = 0$; $[-1, 1]$ for $\sum Q_k = 1$; $[1, -1]$ for $\sum Q_k = -1$. When a π wall forms a closed loop, it has a zero net charge.

The calculated polarization fields indicate that the kinks are demarcations between neighboring subdomains of opposite chirality (Fig. 3e-g). This chirality reversal allows for the continuation of the polarization orientation in the middle plane of the cell, although the polarizations, at two sides of the kink, adopt head-to-head or tail-to-tail configurations at the top or bottom surface, respectively (as indicated by the arrows in Fig. 3d).

We validated these findings in experiments by pressing a cell made with the photoalignment method. The relatively weak anchoring energy leads to a small restoring force, and thus, the twin disclinations are separated to a large distance by the induced shear flow, and take a long time to recover (Supplementary Video 3). By setting the analyzer at 45° or 135° with respect to the polarizer, we could infer the chirality of the enclosed subdomains. As shown in Extended Data Fig. 10, the kinks flip the chirality of subdomains, similar to the spin-up and spin-down process in the one-dimensional Ising model[40].

We further investigated the annihilation dynamics of kinks and antikinks. The $\mathbb{Z}_2$ kinks can move along the $\pi$ wall, exchanging the horizontal arrangements of the twin disclinations. Kinks and antikinks undergo annihilation upon collision (Fig. 3c and Supplementary Video 1). They can be generated in pairs by applying a small pressure to the cell (Supplementary Video 2). When a segment of one disclination traverses over its twin counterpart, a pair of kinks and antikinks is generated. These induced kinks remain stable in thin cells ($h < 3$ μm) but tend to annihilate in thick cells.

The kinks in $N_F$ liquid crystal contrast the Sine-Gordon kinks encountered in solid-state ferroic materials[41], resulting from their continuous and broken translational symmetry, respectively. The Sine-Gordon kink arises from the periodic Peierls potential created by atomic lattices within crystalline materials, and they exhibit an infinite number of ground states[42]. The larger length scales of the $N_F$ liquid crystal make it a model system for experimentally exploring the dynamics of kinks, which have been subjected to extensive theoretical studies[43].

**Topological Constraints on Polar Switching by Electrical Fields**

The topological structures of $\pi$ walls significantly influence polar switching. We explored this process by applying an electrical field along **R**, and gradually increasing its strength (Fig. 4a). The switching of polarizations involves two distinct steps. In the first step, the disclinations on one surface approach each other and are annihilated because of their opposite topological charges, forming a region with a twist of the polarizations across the cell (as illustrated in Fig. 4b-d). There exists an energy barrier impeding the surface disclinations from directly traversing their paired twins. When both disclination lines at one surface are inside their twin lines on the opposite surface, they deform and approach the middle of the domains (see the red dashed line). When one disclination is outside its twin line on the opposite surface, the inside line deforms and approaches the wall of the other domain side (see the blue dashed line). When both disclination lines are outside the twin lines on the opposite surface, they can deform, traverse their paired twins and approach the other wall of the domain (Supplementary Video 4). In the second step, when the electrical field is increased further, the disclinations on the second surface are annihilated, leaving one domain with the polarizations aligned with the electrical field.

We further studied the disclination annihilations by directly applying an electrical field (2 V/mm) that is sufficient to eliminate the disclinations at one confining surface. We found that the chirality of the evolving twist domains plays a dictating role. For example, the twist domains 1 and 2 possess identical chirality that is opposite to that of domains 3 and 4 (Fig. 4e). The approached disclination lines, separating regions of the same chirality, can annihilate (domains 1 and 2; 3 and 4). However, the approached lines, separating regions of opposite twisting chirality, survived (Fig. 4e), as verified by polarized microscopic imaging (Fig. 4f). These disclinations unite and form a $2\pi$ wall, as suggested in a prior study[32].

**Discussion**

These findings beget several important implications and questions for future explorations. First, walls, other than $\pi$ walls, are expected in the uniformly aligned cells (Extended Data Fig. 11), separating (i) unidirectional and $\pi$-twisted domains, (ii) left- and right-handed $\pi$-twisted domains, and (iii) two $\pi$-twisted domains of identical chirality but opposite polarizations at the two confining substrates. In the first case, the domain wall consists of a single surface disclination at one substrate, and can be referred to as a half $\pi$ wall[5]. The second case includes two possibilities: 1) when the polarizations of the two domains are identical at the confining substrates, the wall is a disclination in the middle of the cell and is referred to as a $2\pi$ wall[32]; and 2) when the polarizations of the two domains are opposite at the substrates, the wall may feature two paired surface disclinations and is referred as an asynchronous-twist (AT) $\pi$ wall. The last case is equally intriguing, as the polarizations seem opposite throughout the cell. This wall may consist of two surface disclinations and is referred to as a synchronous twist (ST) $\pi$ wall.

Second, the arrangements of the polarizations around surface disclinations generate a bound charge of density $\rho_d = |\mathbf{P}|/r$, where $r$ is the radius of the core. Although a prior study suggested that the tilt of $\mathbf{P}$ yields a gigantic depolarization field[44], it becomes possible by (i) the depression of $\mathbf{P}$ and (ii) the segregation of free charges at the defect core, screening the induced bound charge.

The orientation of the tilted polarizations should depend on the accumulated charges (either positive or negative), or the flexoelectric coupling effect. The segregation of free charges may lead to an intriguing property: the domain walls in $N_F$ liquid crystals could be electrically conductive.

**References**

1. Debye, P. Einige resultate einer kinetischen theorie der isolatoren. *Physik Z.* **13**, 97 (1912).

2. Born, M. Über anisotrope Flüssigkeiten. Versuch einer Theorie der flüssigen Kristalle und des elektrischen Kerr-Effekts in Flüssigkeiten. *Sitzungsber. Preuss. Akad Wiss.* 614–650 (1916).

3. Nishikawa, H. *et al.* A Fluid Liquid-Crystal Material with Highly Polar Order. *Advanced Materials* **29**, 1702354 (2017).

4. Mandle, R. J., Cowling, S. J. & Goodby, J. W. A nematic to nematic transformation exhibited by a rod-like liquid crystal. *Phys. Chem. Chem. Phys.* **19**, 11429–11435 (2017).

5. Chen, X. *et al.* First-principles experimental demonstration of ferroelectricity in a thermotropic nematic liquid crystal: Polar domains and striking electro-optics. *Proc. Natl. Acad. Sci. U.S.A.* **117**, 14021–14031 (2020).

6. Feng, C. *et al.* Electrically Tunable Reflection Color of Chiral Ferroelectric Nematic Liquid Crystals. *Advanced Optical Materials* **9**, 2101230 (2021).

7. Barboza, R. *et al.* Explosive electrostatic instability of ferroelectric liquid droplets on ferroelectric solid surfaces. *Proc. Natl. Acad. Sci. U.S.A.* **119**, e2207858119 (2022).

8. Marni, S., Nava, G., Barboza, R., Bellini, T. G. & Lucchetti, L. Walking Ferroelectric Liquid Droplets with Light. *Advanced Materials* **35**, 2212067 (2023).

9. Zhao, X. *et al.* Nontrivial phase matching in helielectric polarization helices: Universal phase matching theory, validation, and electric switching. *Proceedings of the National Academy of Sciences* **119**, e2205636119 (2022).

10. Sultanov, V. *et al.* Tunable entangled photon-pair generation in a liquid crystal. *Nature* 1–6 (2024) doi:10.1038/s41586-024-07543-5.

11. Yang, J. *et al.* Spontaneous electric-polarization topology in confined ferroelectric nematics. *Nat Commun* **13**, 7806 (2022).

12. Yang, J., Zou, Y., Li, J., Huang, M. & Aya, S. Flexoelectricity-driven toroidal polar topology in liquid-matter helielectrics. *Nat. Phys.* (2024) doi:10.1038/s41567-024-02439-7.

13. Kumari, P., Basnet, B., Lavrentovich, M. O. & Lavrentovich, O. D. Chiral ground states of ferroelectric liquid crystals. *Science* **383**, 1364–1368 (2024).

14. Karcz, J. *et al.* Spontaneous chiral symmetry breaking in polar fluid–heliconical ferroelectric nematic phase. *Science* **384**, 1096–1099 (2024).

15. Catalan, G., Seidel, J., Ramesh, R. & Scott, J. F. Domain wall nanoelectronics. *Rev. Mod. Phys.* **84**, 119–156 (2012).

16. Meier, D. & Selbach, S. M. Ferroelectric domain walls for nanotechnology. *Nat Rev Mater* **7**, 157–173 (2021).

17. Zhirnov, V. A. A contribution to the theory of domain walls in ferroelectrics. *J. Exptl. Theoret. Phys* **35**, 1175–1180 (1959).

18. Wei, X.-K. *et al.* Néel-like domain walls in ferroelectric Pb(Zr,Ti)O3 single crystals. *Nat Commun* **7**, 12385 (2016).

19. Cherifi-Hertel, S. *et al.* Non-Ising and chiral ferroelectric domain walls revealed by nonlinear optical microscopy. *Nat Commun* **8**, 15768 (2017).

20. Weymann, C. *et al.* Non-Ising domain walls in c -phase ferroelectric lead titanate thin films. *Phys. Rev. B* **106**, L241404 (2022).

21. Zhou, J., Xia, R., Huang, M. & Aya, S. Stereoisomer effect on ferroelectric nematics: stabilization and phase behavior diversification. *J. Mater. Chem. C* **10**, 8762–8766 (2022).

22. Mrukiewicz, M., Perkowski, P., Karcz, J. & Kula, P. Ferroelectricity in a nematic liquid crystal under a direct current electric field. *Phys. Chem. Chem. Phys.* **25**, 13061–13071 (2023).

23. Stepanafas, G. *et al.* Ferroelectric nematogens containing a methylthio group. *Mater. Adv.* 10.1039.D3MA00446E (2023) doi:10.1039/D3MA00446E.

24. Basnet, B. *et al.* Soliton walls paired by polar surface interactions in a ferroelectric nematic liquid crystal. *Nat Commun* **13**, 3932 (2022).

25. Li, J. *et al.* Development of ferroelectric nematic fluids with giant-$\varepsilon$ dielectricity and nonlinear optical properties. *Sci. Adv.* **7**, eabf5047 (2021).

26. Sebastián, N. *et al.* Ferroelectric-Ferroelastic Phase Transition in a Nematic Liquid Crystal. *Phys. Rev. Lett.* **124**, 037801 (2020).

27. Lavrentovich, O. D. Ferroelectric nematic liquid crystal, a century in waiting. *Proc. Natl. Acad. Sci. U.S.A.* **117**, 14629–14631 (2020).

28. Kittel, C. Theory of the Structure of Ferromagnetic Domains in Films and Small Particles. *Phys. Rev.* **70**, 965–971 (1946).

29. Mertelj, A. *et al.* Splay Nematic Phase. *Phys. Rev. X* **8**, 041025 (2018).

30. Sebastián, N. *et al.* Polarization patterning in ferroelectric nematic liquids via flexoelectric coupling. *Nat Commun* **14**, 3029 (2023).

31. Caimi, F. *et al.* Surface alignment of ferroelectric nematic liquid crystals. *Soft Matter* **17**, 8130–8139 (2021).

32. Chen, X. *et al.* Polar in-plane surface orientation of a ferroelectric nematic liquid crystal: Polar monodomains and twisted state electro-optics. *Proc. Natl. Acad. Sci. U.S.A.* **118**, e2104092118 (2021).

33. Han, Y., Yin, J., Hu, Y., Majumdar, A. & Zhang, L. Solution landscapes of the simplified Ericksen–Leslie model and its comparisonwith the reduced Landau–deGennes model. *Proc. R. Soc. A.* **477**, 20210458 (2021).

34. Ericksen, J. L. Liquid crystals with variable degree of orientation. *Arch. Rational Mech. Anal.* **113**, 97–120 (1991).

35. Nataf, G. F. *et al.* Domain-wall engineering and topological defects in ferroelectric and ferroelastic materials. *Nat Rev Phys* **2**, 634–648 (2020).

36. Smalyukh, I. I. Review: knots and other new topological effects in liquid crystals and colloids. *Rep. Prog. Phys.* **83**, 106601 (2020).

37. Mermin, N. D. The topological theory of defects in ordered media. *Rev. Mod. Phys.* **51**, 591–648 (1979).

38. Vitek, V. & Kléman, M. Surface disclinations in nematic liquid crystals. *J. Phys. France* **36**, 59–67 (1975).

39. Vachaspati, T. *Kinks and Domain Walls: An Introduction to Classical and Quantum Solitons*. (Cambridge University Press, 2006). doi:10.1017/CBO9780511535192.

40. Ising, E. Beitrag zur Theorie des Ferromagnetismus. *Z. Physik* **31**, 253–258 (1925).

41. Buijnsters, F. J., Fasolino, A. & Katsnelson, M. I. Motion of Domain Walls and the Dynamics of Kinks in the Magnetic Peierls Potential. *Phys. Rev. Lett.* **113**, 217202 (2014).

42. Novoselov, K. S., Geim, A. K., Dubonos, S. V., Hill, E. W. & Grigorieva, I. V. Subatomic movements of a domain wall in the Peierls potential. *Nature* **426**, 812–816 (2003).

43. Christov, I. C. *et al.* Kink-Kink and Kink-Antikink Interactions with Long-Range Tails. *Phys. Rev. Lett.* **122**, 171601 (2019).

44. Caimi, F. *et al.* Fluid superscreening and polarization following in confined ferroelectric nematics. *Nat. Phys.* (2023) doi:10.1038/s41567-023-02150-z.

45. Yeh, P. & Gu, C. *Optics of Liquid Crystal Displays*. (Wiley, New York, 1999).

46. Shribak, M. & Oldenbourg, R. Techniques for fast and sensitive measurements of two-dimensional birefringence distributions. *Appl. Opt., AO* **42**, 3009–3017 (2003).

47. Okano, K. Electroslatie Contribution to the Distortion Free Energy Density of Ferroelectric Liquid Crystals. *Jpn. J. Appl. Phys.* **25**, L846 (1986).

48. Lee, J.-B., Pelcovits, R. A. & Meyer, R. B. Role of electrostatics in the texture of islands in free-standing ferroelectric liquid crystal films. *Phys. Rev. E* **75**, 051701 (2007).

49. Poy, G. & Žumer, S. Ray-based optical visualisation of complex birefringent structures including energy transport. *Soft Matter* **15**, 3659–3670 (2019).

50. Poy, G. & Žumer, S. Physics-based multistep beam propagation in inhomogeneous birefringent media. *Opt. Express* **28**, 24327 (2020).

**Methods**

**Mechanically rubbed cells**

The glass substrates were ultrasonically cleaned in deionized water and dried by blowing nitrogen gas. We spin-coated a 10 wt % dimethylformamide (DMF) solution of polyimide (PI-2555) at ~4000 rpm on these cleaned substrates and then baked them at 200 ℃ for 90 minutes. The PI films were rubbed by using a home-built setup (Extended Data Fig. 1a). The roller has a diameter of ~10 mm and is wrapped by a flannel cloth of 2 mm thickness. The PI layer was unidirectionally rubbed by rotating the roller with an angular velocity of approximately 30 rad/s for 5 seconds. The rubbing process induced nano-sized grooves, as shown in the scanning electron microscopy (SEM) image in Extended Data Fig.1d. The cells were assembled by using two rubbed substrates; the rubbing directions of the two substrates are parallel or antiparallel. The cell thickness was controlled by using silica beads as spacers, ranging from 1 μm to 10 μm. The DMF and PI-2555 were purchased from Sigma-Aldrich and HD Microsystems, respectively, and used without further purification.

**Photoaligned Cells**

We also made uniform-aligned cells by using photoalignment methods. The glass substrates were cleaned as described above and treated with ultraviolet light and ozone in a plasma cleaning equipment for 10 minutes. A layer of azobenzene dye SD1 (0.5 wt% in DMF) was spin-coated on the glass substrates at a speed of 3000 rpm for 30 seconds. The liquid crystal cells were assembled with the SD1-coated surfaces facing each other. The assembled cells were then exposed to linearly polarized light (470 nm, 760 mW) for 10 minutes to induce a uniform alignment.

To apply an in-plane electric field, we used glass substrates prepatterned with ITO stripe electrodes of 1 mm spacing to fabricating liquid crystal cells.

**Polarized, Polscope and SHG optical microscopies**

The polarized microscopy images were recorded using a Leica Microsystem (DM2700 M) with a CMOS camera (DMC5400).

The subdomain features an π twist of polarizations along the cell normal. Its chirality can be inferred based on the Jones matrix method. Let Γ be the phase retardation of the cell when it is untwisted: $\Gamma = 2\pi \Delta n h/\lambda$, where $h$, $\Delta n$, and $\lambda$ represent the thickness of the cell, the birefringence, and the wavelength of light, respectively. We assumed that the twist is linear, and the total twist angle is $\phi$. Under the polarized microscopy, the transmitted intensity can be written as[45]:

$$I = \cos^2(\phi - \Phi_{exit} + \Phi_{ent}) + \sin^2 X \sin2(\phi - \Phi_{exit})\sin2\Phi_{ent}$$
$$+ \frac{\phi}{2X}\sin2X\sin2(\phi - \Phi_{exit} + \Phi_{ent}) - \phi^2\frac{\sin^2 X}{X^2}\cos2(\phi \quad (1)$$
$$- \Phi_{exit})\cos2\Phi_{ent}$$

where $X = \sqrt{\phi^2 + (\Gamma/2)^2}$, $\Phi_{ent}$ and $\Phi_{exit}$ are the orientation angles of transmission axes of the polarizer and analyser, respectively.

We plotted the transmitted light intensity ($I$) as a function of wavelength for $\Phi_{exit} = 45°$ (Extended Data Fig. 6a) and for $\Phi_{exit} = 135°$ (Extended Data Fig. 6b). Under uncrossed polarizers, twist domains of opposite chirality demonstrate contradictory trends in their optical properties. Specifically, when $\Phi_{exit} = 45°$, transmitted intensity of the left-handed twist domain features a minimum point at $\lambda \approx 490$ nm, while the transmitted intensity of the right-handed twist domain shows a maximum value. When $\Phi_{exit} = 135°$, their transmitted intensity curves (colours under polarized microscopy) exchange.

The PolScope measurement was performed using a home-built system. The incident beam has a wavelength of 544 nm and its polarity was tuned with a system of polarizers followed by quarter-wave plate. Image acquisition was done using the Leica Microsystem (DM2700 M, 20× objective) with a CMOS camera (FL 9BW, SONY IMX533CLK-D). A series of polarization settings were used for measuring two-dimensional distributions of linear birefringence. Director patterns were calculated based on an image processing algorithm detailed in Ref.[46].

SHG imaging was performed using a two-photon optical microscopy (LSM 980, Carl Zeiss). The probing light has a wavelength of 800 nm and its power was chosen below the damage threshold of the sample. The incident beam was circularly polarized and directed perpendicular into the cell. The second harmonic emission at 400 nm was simultaneously collected by the detector.

**Measurements of the domain widths and twin disclination spacings**

To measure the spacing between twin disclinations, we set the crossed polarizers at 45° with respect to the rubbing direction; hence the whole ferroelectric domains appear bright. Next, we adjusted the focus plane of the microscope to clearly observe the twin lines and took images of the samples. π walls consist of one bright and one dark lines. We plotted the transmitted intensity profiles across a domain wall, which features a minimum and a maximum point (Extended Data Fig.5. The distances between these two points were measured as the line spacing. Each measured spacing represents the average over 10 intensity profiles.

To measure the average domain widths, we drew a line perpendicular to the rubbing direction and counted the total number of unidirectional domains intersected. The domain widths are the lengths of the line divided by the number of domains. We averaged the measurements over 5-10 regions for each cell.

**Numerical calculations**

To calculate the equilibrium polarization fields and the total free energy of $\pi$ walls, we adopted numerical simulations based on the modified Ericksen-Leslie model[33,34]:

$$F = \int_V (f_{\text{ela}} + f_{\text{grad}} + f_{\text{pote}}) \, dV + \int_S f_{\text{surf}} \, dS. \tag{2}$$

The elastic energy density $f_{\text{ela}}$ associated with the deformation of directors is given by:

$$f_{\text{ela}} = \frac{p^2}{2} K_{11} (\nabla \cdot \mathbf{n})^2 + \frac{p^2}{2} K_{22} (\mathbf{n} \cdot \nabla \times \mathbf{n})^2 + \frac{p^2}{2} K_{33} |\mathbf{n} \times \nabla \times \mathbf{n}|^2, \tag{3}$$

where $K_{11}$, $K_{22}$, and $K_{33}$ are the elastic constants for the splay, twist, and bend deformations, respectively; $\mathbf{n}$ is a unit vector representing director field that is parallel to the polarization: $\boldsymbol{P} \equiv p|P_0|\mathbf{n}$, where $P_0$ is a constant, expressing the saturation value of the spontaneous polarization; $p$ is the polar order parameter. We took into account the energy increase due to a gradient of $p$ by: $f_{\text{grad}} = \frac{b}{2} |\nabla p|^2$, where $b$ is an elastic constant.

We adopted a Ginzburg-Landau form of bulk free energy density:

$$f_{\text{pote}} = \frac{1}{4} \eta_0 (p^2 - p_0^2)^2 \tag{4}$$

where $\eta_0$ is a material constant. The bulk free energy enforces a preferred polar order $p_0$ in the bulk. The potential has two minima at $\pm p_0$, as required by the reflection symmetry. The surface energy is modelled through an expression: $f_{\text{surf}} = \frac{1}{2} W (\mathbf{n} - \mathbf{n}_s)^2$, where $\mathbf{n}_s$ is the surface preferred director and $W$ represents the anchoring strength.

Splay deformations of polarizations lead to bound changes, as known from the Gauss law: $\rho = -\nabla \cdot \boldsymbol{P}$, which increases the free energy because of the depolarization fields[47,48]. This effect can be taken into account by: $K_{11} = K_0(1 + \lambda_D^2/\xi_P^2)$, where $K_0$ is the elastic modulus measured for paraelectric liquid crystals; $\lambda_D$ is the Debye length and $\xi_P$ represents the polarization penetration length.

As detailed in Ref. [24], the flexoelectricity coupling leads effectively to an increase of the elastic constant $K_{11}$ in the $N_F$ phase. To consider the contribution of electrostatic energy and the flexoelectricity, we employed a relatively large $K_{11}$ in the simulation. Unless otherwise specified, the following parameters are used: $K_{11} = 6$ pN, $K_{22} = 1$ pN, $K_{33} = 2$ pN, $p_0 = 0.9$, $b = 1$ pN, $\eta_0 = 0.5 \times 10^6$ Jm$^{-3}$ and $W = 5 \times 10^{-2}$ Jm$^{-2}$.

We set twin lines at two confining surfaces as the Bloch type and horizontally separated. The director field at confining surfaces is $\mathbf{n}_s = (0, \cos\theta, \sin\theta)$, with $\theta = \frac{\pi}{2}\tanh[(x - X_t)/\lambda]$, where $X_t$ and $X_b$ represent the locations of lines at the top and bottom surface, respectively. We set $\lambda = 500$ nm and the width of the line becomes $2\lambda = 1$ μm, which is close to the value observed in experiments. We used periodical boundary conditions in the lateral directions. We used the $\mathbf{n}$ fields derived from the polarized microscopy and SHG images, with random disturbance as initial conditions. Minimization of the energy was carried out according to the Euler–Lagrange equation. The equation was solved by a finite difference method on a cubic mesh of spacing 50 nm. We computed corresponding optical textures under crossed polarizers based on simulated director fields using a standard simulation procedure detailed in Ref [49,50]. Visualization of the director field was performed in ParaView, an open-source freeware.

**Data availability** The data generated and analyzed during the current study are available from the corresponding authors on reasonable request.

**Acknowledgments** We thank Oleg D. Lavrentovich, Jonathan V. Selinger, Robin L. B. Selinger, Jiangyu Li, Junmin Liu and Lang Chen for the helpful discussions.

**Funding** Financial support by the National Key Research and Development Program of China via grant 2022YFA1405000, and the National Nature Science Foundation of China via grant 6210030761 and 12204226, and the Guangzhou Basic and Applied Basic Research Foundation via grant 2024B1515040023 are acknowledged.

**Author contributions** Conceptualization: Q.W., S.Y.; Methodology: S.Y., Z. H., C.Z., Z.M., M. H., S. A.; Investigation: S.Y., Z.H., C.Z., Z.M.; Visualization: S.Y.; Funding acquisition: Q.W., S. A., R.Z., M. J.; Project administration: Q.W., M.J.; Supervision: Q.W., R.Z.; Writing−original draft: S.Y.; Writing − review & editing: Q.W., S.Y., R.Z, S. A.

**Competing interests** Authors declare that they have no competing interests.

**Figures and Captions**

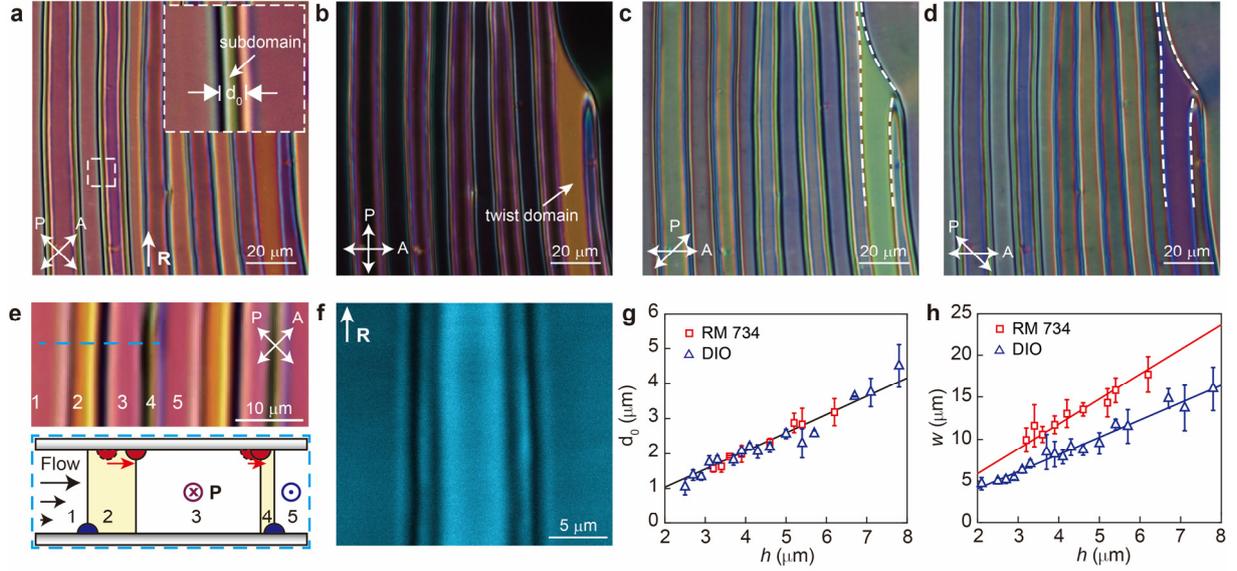

**Fig.1|Experimental results for π walls in $N_F$ liquid crystals.** (**a-d**) Polarized microscopy images of π walls in a rubbed cell of thickness $h = 5.1$ μm. The material DIO is used. The crossed polarizers are at 45° (**a**) or parallel/perpendicular (**b**) to the rubbing direction. The polarizer is rotated 45° clockwise (**c**) and anticlockwise (**d**) with respect to the analyzer. (**e**) Separation of twin lines of π walls by shear. A schematic cross section along the dashed line is shown in the bottom panel. The yellow box and red arrow represent subdomains and motions of lines at the top substrate. (**f**) SHG image obtained by a circularly polarized incident light. (**g**) Horizontal separation $d_0$ of twin lines versus the cell thickness. (**h**) Domain width as a function of the cell thickness. The markers and solid lines in (**g**) and (**h**) represent the experimental measurements and linear fits, respectively; the error bars indicate the standard deviation.

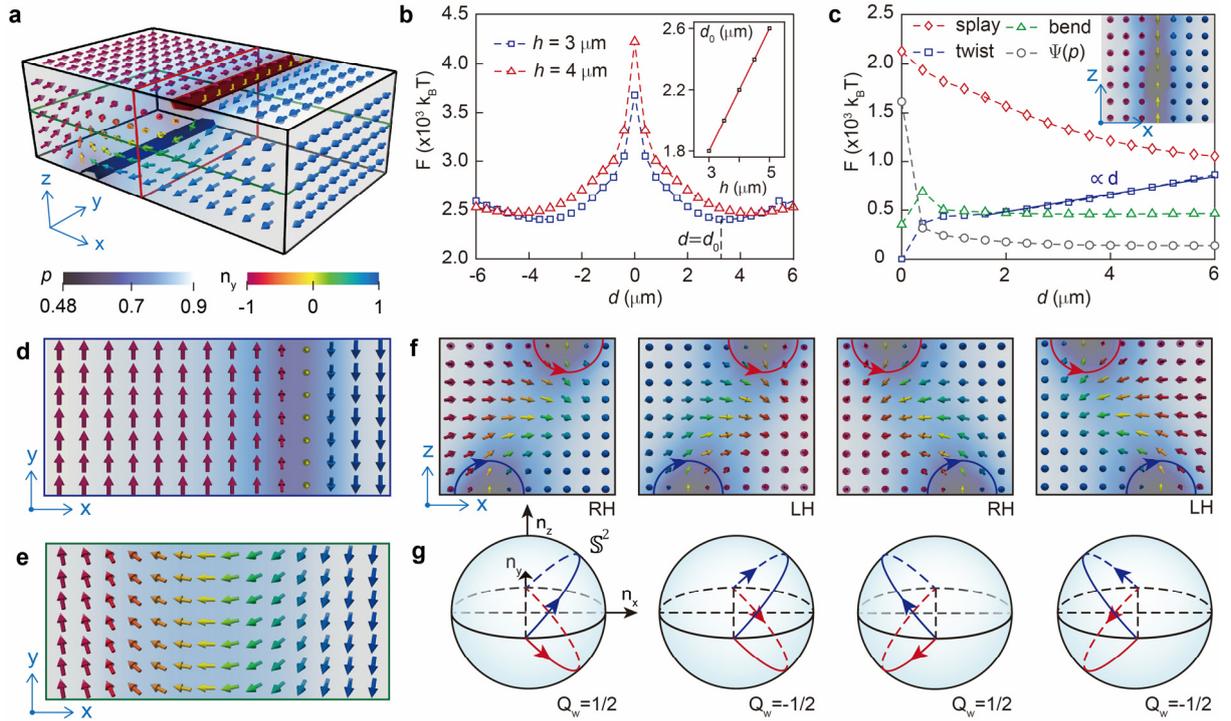

**Fig.2|Topological structures of the π wall.** (**a**) Simulated structures of a π wall. An isosurface where $p = 0.65$ is used to visualize the disclinations. The disclinations on the top and bottom surfaces are colored red and blue, respectively. (**b**) Free energy cost of a π wall versus $d$. The insert shows the calculated optimal separation $d_0$ versus cell thickness. (**c**) Energy costs of different terms versus $d$. The insert shows the simulated polarization field when two lines overlap. Simulated polarization fields near the top surface (**d**) and in the middle plane of the cell (**e**). (**f**) Exemplary polarization fields of π walls with different chirality. (**g**) Topological schemes of the domain walls; along the red and blue lines circling disclinations shown in (**f**), the polarization rotates by π, thus covering a half-circle.

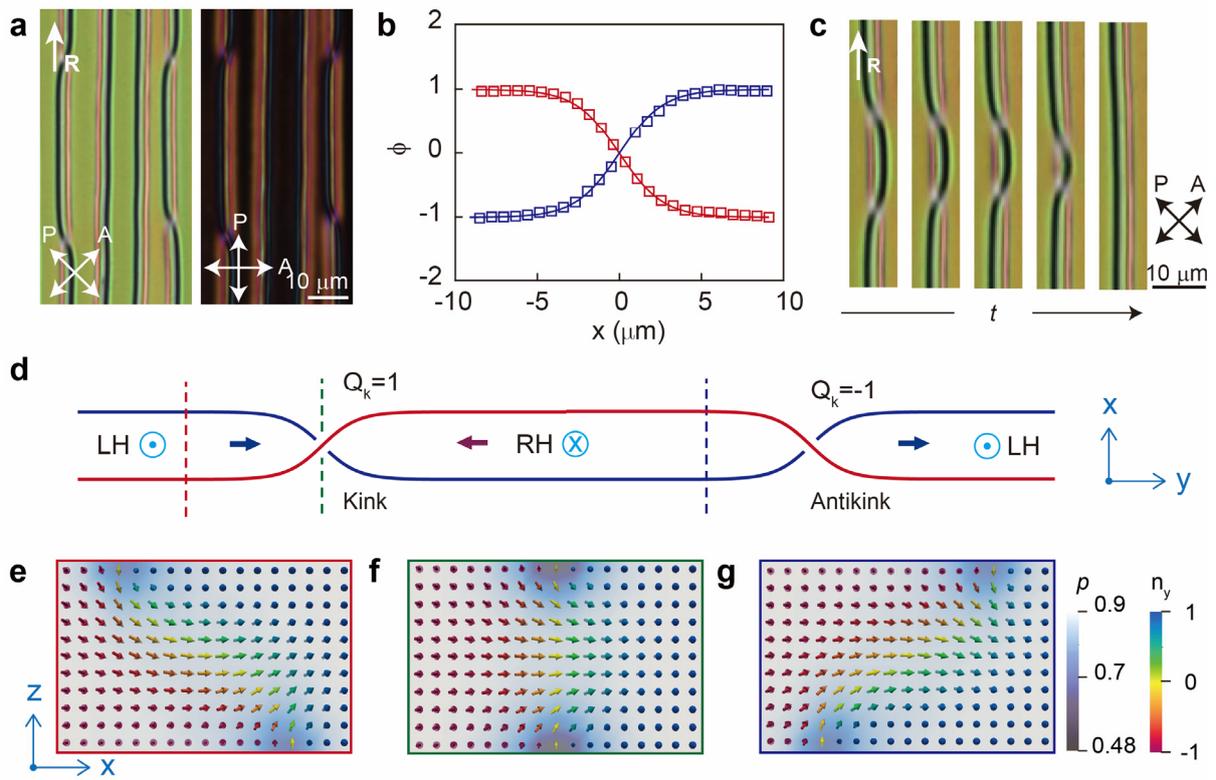

**Fig.3|Kinks and antikinks along a π wall.** (**a**) Polarized microscopy images of kinks and antikinks along a domain wall. The thickness of the rubbed cell is 4.4 μm. The liquid crystal material DIO is used. (**b**) Profile of kinks. Solid lines are numerical fits with $\xi$ as the fitting parameter. (**c**) Time snapshots showing the annihilation of kinks and antikinks. (**d**) Illustration showing the top view of kinks and antikinks. Disclinations in the top and bottom substrates are represented by red and blue lines, respectively. The polarizations of the subdomain at the top surface are represented by red and blue arrows. (**e-g**) Simulated polarization fields of the cross sections shown in (**d**).

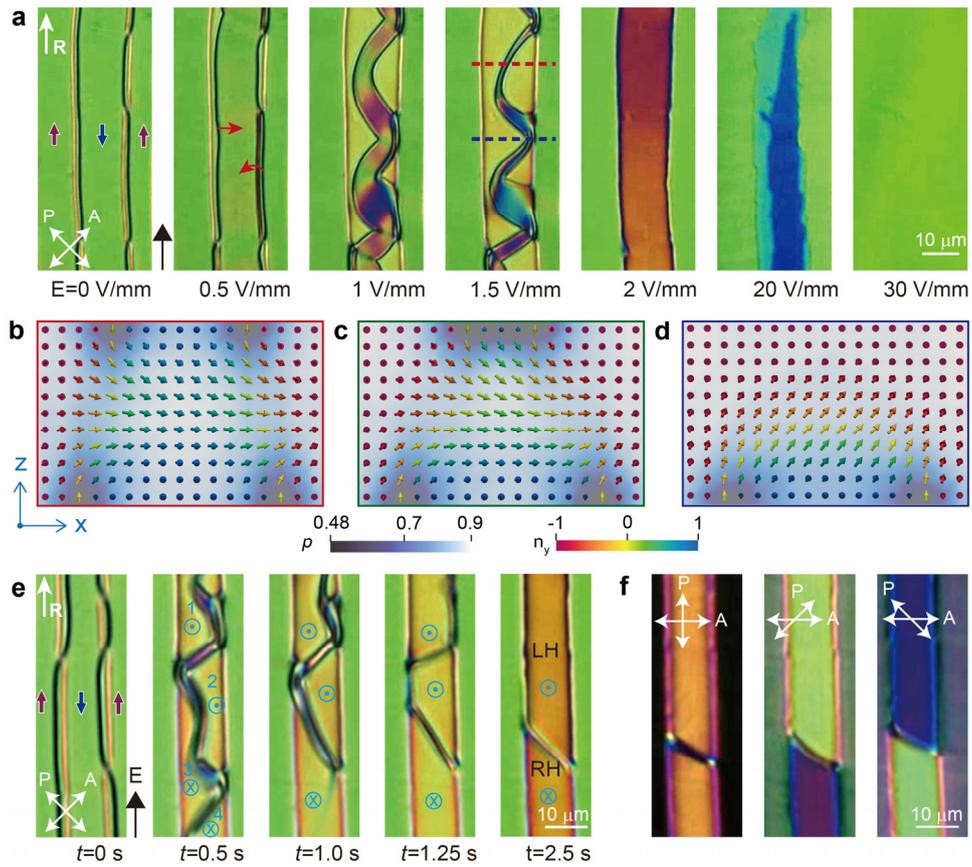

**Fig.4|Two-step electrical switching.** (**a**) Polarized microscopy images showing domain switching under an in-plane electric field. The magnitude of the electric field increased from 0 to 30 V/mm. (**b-d**) Schematic illustrations showing the polarization fields in the first step; two lines on the same substrates came into contact (**c**) and were annihilated, leaving a twist domain (**d**). (**e**) Polarized microscopy textures showing the formation of two twist domains of opposite chirality, under an in-plane electric field of 2 V/mm. (**f**) Polarized microscopy textures of the evolved domains under crossed and uncrossed polarizers. The material RM734 is used; the cell thickness $h = 3.3$ μm.

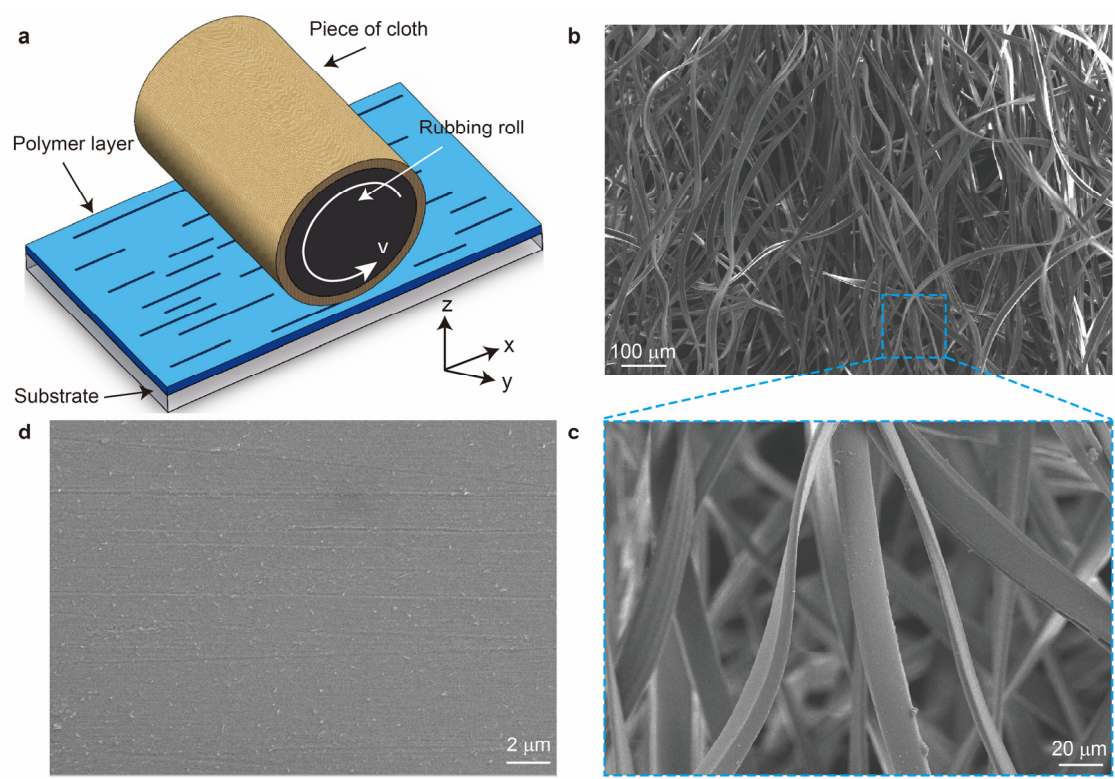

**Extended Data Fig.1|Experimental details of the rubbing method** (**a**) Schematic experimental setup for rubbing of PI films coated on glass substrates. (**b**) SEM image of the flannel used for rubbing. (**c**) An enlarged area of (**b**). (**d**) SEM image of a rubbed PI film.

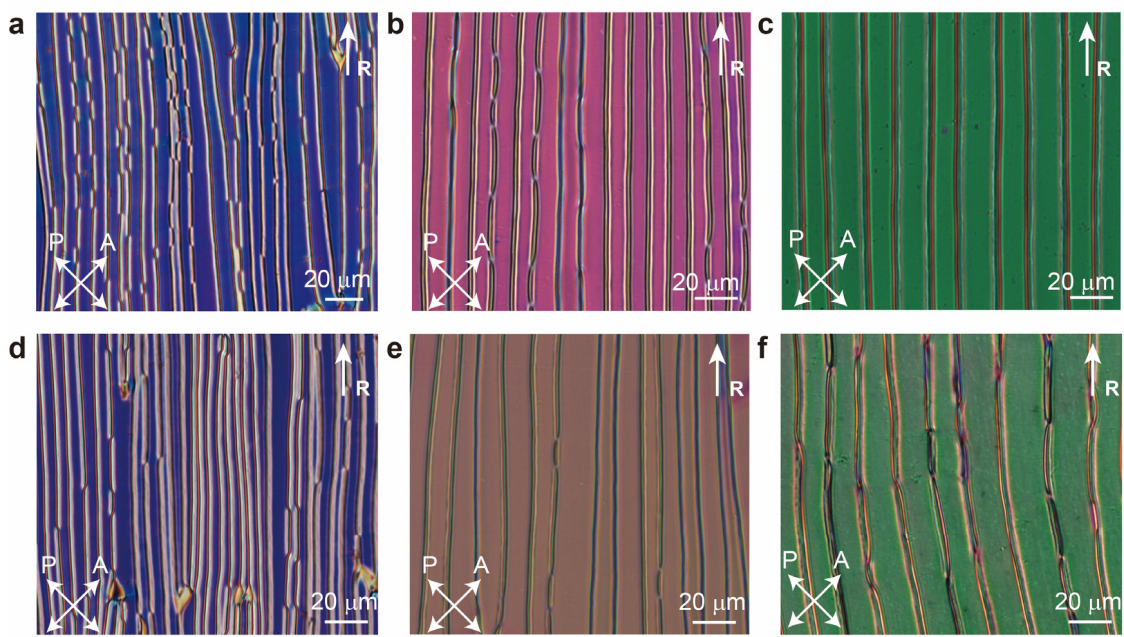

**Extended Data Fig.2|Polarized optical microscopy images of domains and domain walls in rubbed cells.** The rubbing directions at two substrates of the cell are parallel. (**a-c**) The cells are filled with DIO, and the cell thicknesses are: 3.0 μm (**a**); 5.3 μm (**b**), and 7.0 μm (**c**). (**d-f**) The cells are filled with RM734, and the cell thicknesses are: 2.4 μm (**d**), 4.0 μm (**e**), and 5.5 μm (**f**). Temperatures were kept at 60℃ and 120℃ for DIO and RM734, respectively.

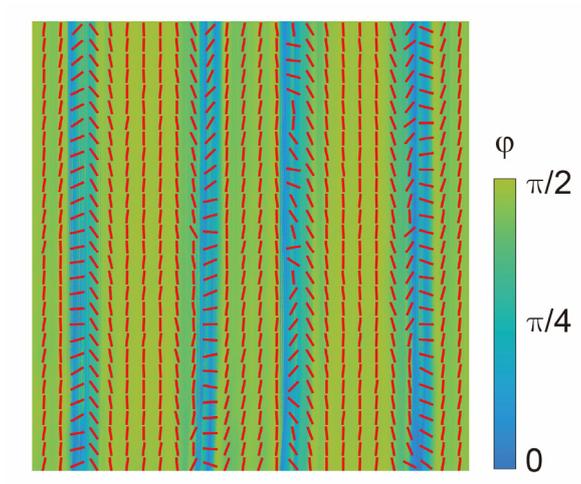

**Extended Data Fig.3|Director field measured by the PolScope technique.** The background is colored based on the azimuthal angle ($\varphi$) of directors.

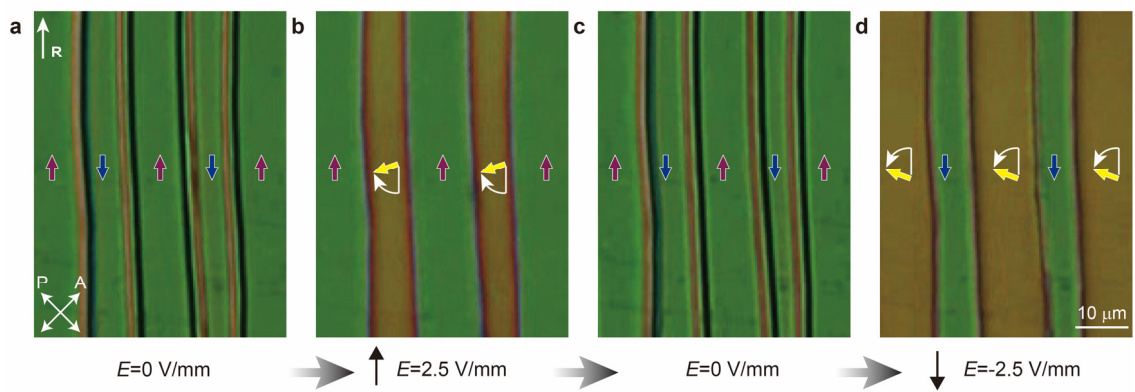

**Extended Data Fig.4|Evidence of π Walls.** Polarized microscopy images of the ferroelectric stripes taken sequentially: (**a**) at zero applied electric field, (**b**) after applying an electric field along the rubbing direction, (**c**) after switching off the electric field, (**d**) after applying an electric field in the opposite direction. The cell thickness is 3 μm; the liquid crystal material RM734 is used. These results demonstrate that two adjacent domains have opposite polarizations and that the domain walls are π walls.

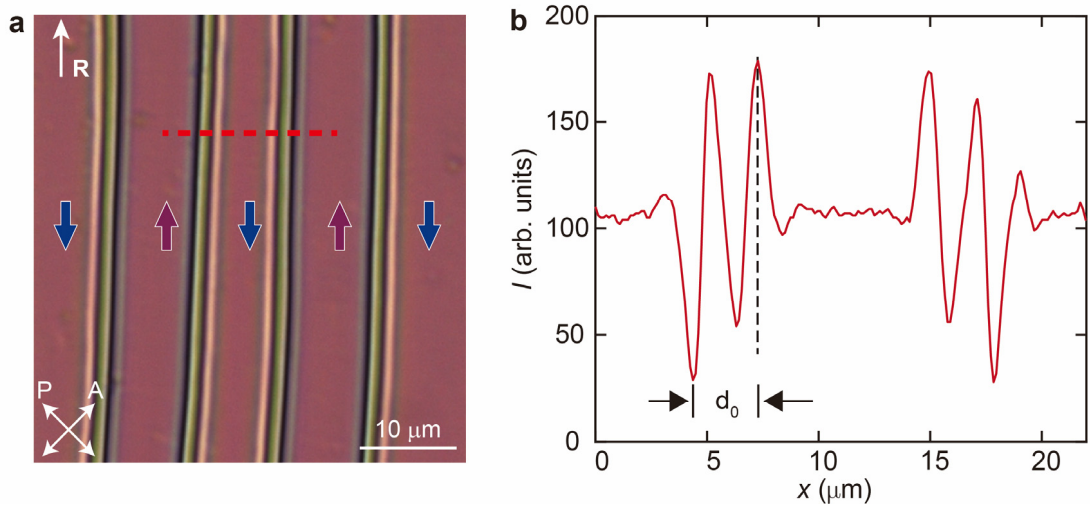

**Extended Data Fig.5 | Schematic way for measuring separations between twin disclinations.** (**a**) Polarized optical microscopy texture when the crossed polarizers are at 45 degrees with respect to the rubbing direction. (**b**) Transmitted intensity profile along the red dashed line in (**c**), which crosses two neighboring π walls. The thickness of the cell is $h = 5.1$ μm and the liquid crystal material is DIO.

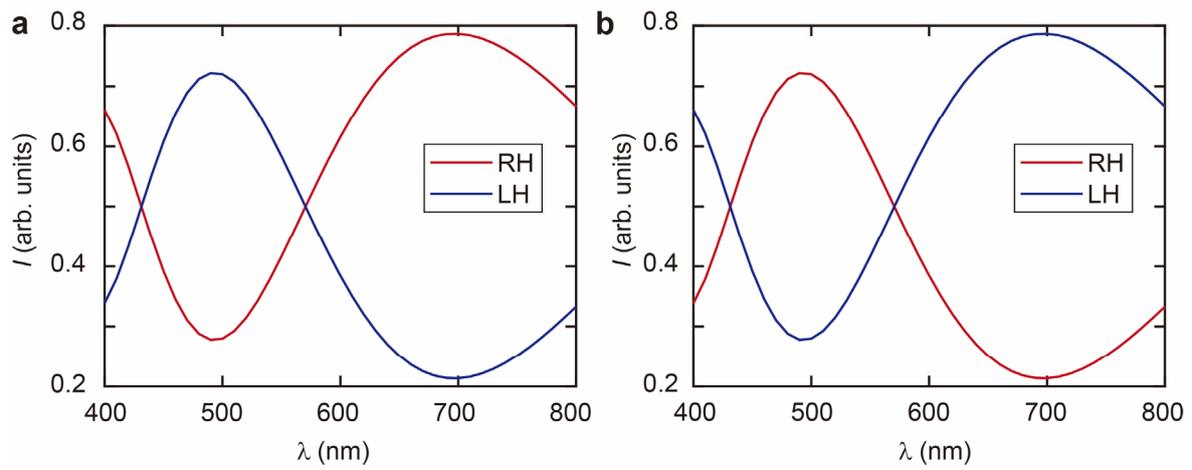

**Extended Data Fig.6|Transmitted intensity of twist domains versus wavelength.** The transmitted intensity was calculated based on Eq. S1, where $h = 5.1$ μm, $\Delta n = 0.19$, $\phi = \pm\pi$, $\Phi_{ent} = 0°$, $\Phi_{exit} = 45°$ (**a**), and $\Phi_{exit} = 135°$ (**b**).

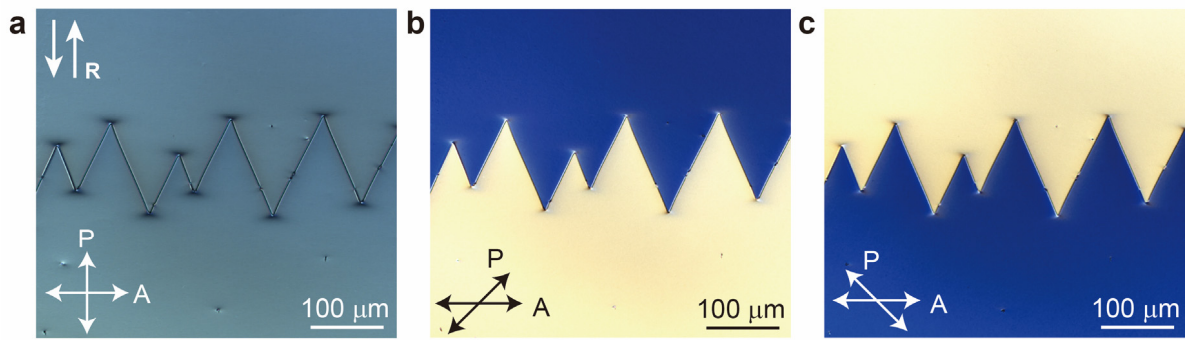

**Extended Data Fig.7|Polarized optical microscopy images of domains and domain walls in rubbed cells.** The rubbing directions at two substrates of the cell are antiparallel. The crossed polarizers are parallel/perpendicular (**a**) to the rubbing direction. The polarizer is rotated 45° clockwise (**b**) and anticlockwise (**c**) with respect to the analyzer. The cell thickness $h = 4.6$ μm; the material DIO is used.

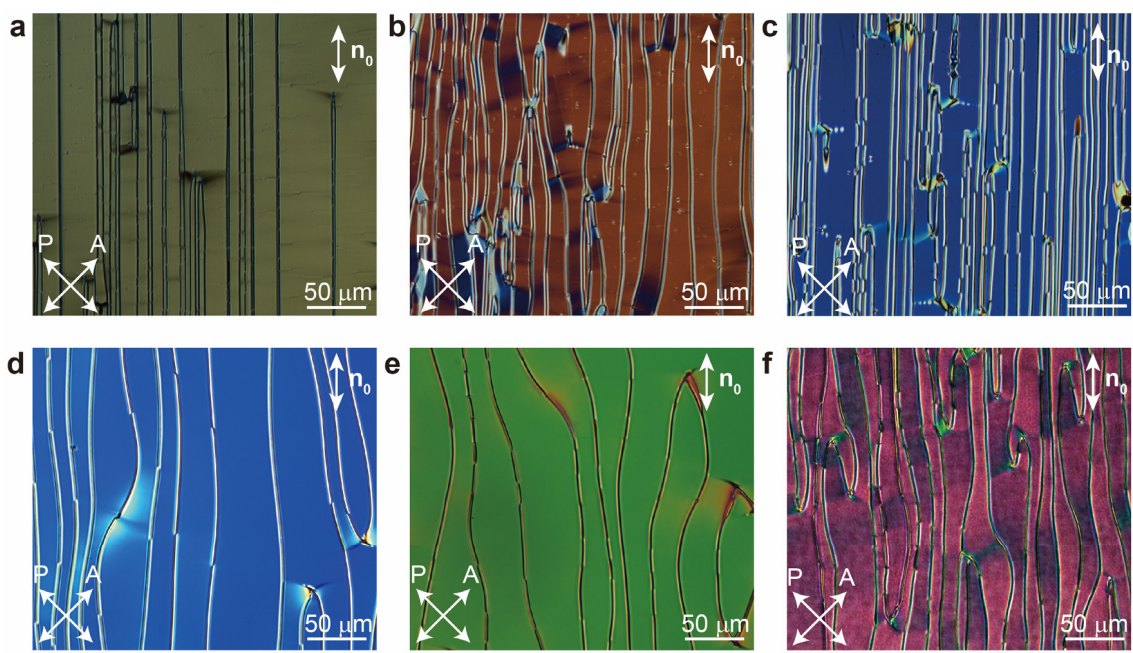

**Extended Data Fig.8|Polarized optical microscopy images of photo-aligned cells.** (**a-c**) Cells are filled with DIO, and cell thicknesses are: 1.8μm (**a**), 2.4 μm (**b**) and 3.1 μm (**c**). (**d-f**) Cells are filled with RM734, and cell thicknesses are: 2.6 μm (**d**), 3.5 μm (**e**) and 4.2 μm (**f**). Temperatures were kept at 60°C and 120°C for DIO and RM734, respectively.

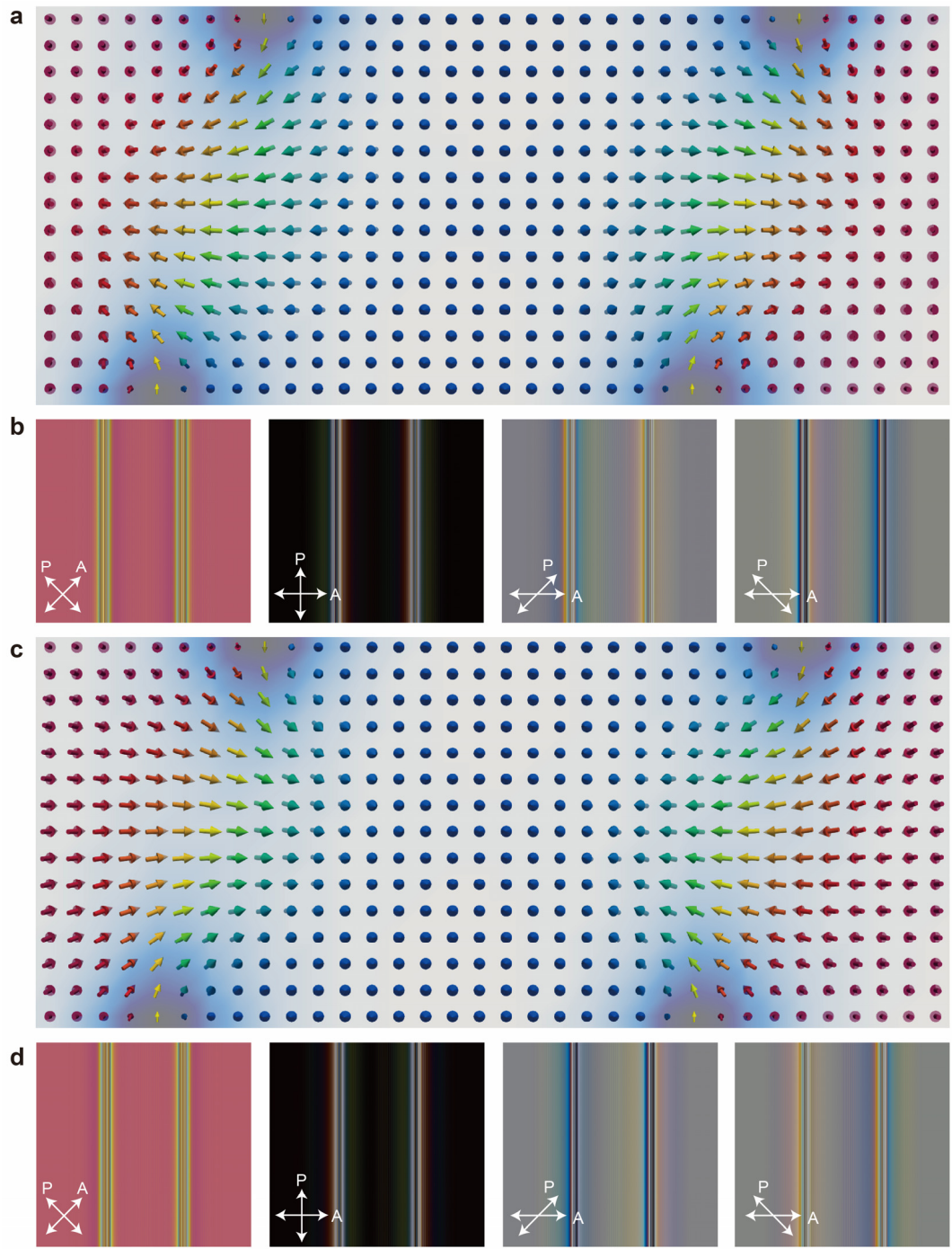

**Extended Data Fig.9|Calculated polarization fields and their corresponding polarized optical images.** (**a**, **b**) Calculated polarization field (**a**) and its corresponding polarized optical textures (**b**) for two adjacent $\pi$ walls with left-handed twisted subdomains. (**c**, **d**) Calculated polarization field (**c**) and its corresponding polarized optical textures (**d**) for two adjacent $\pi$ walls with right-handed twisted subdomains. The cell thickness is $h = 5.1$ μm and the birefringence is $\Delta n = 0.19$.

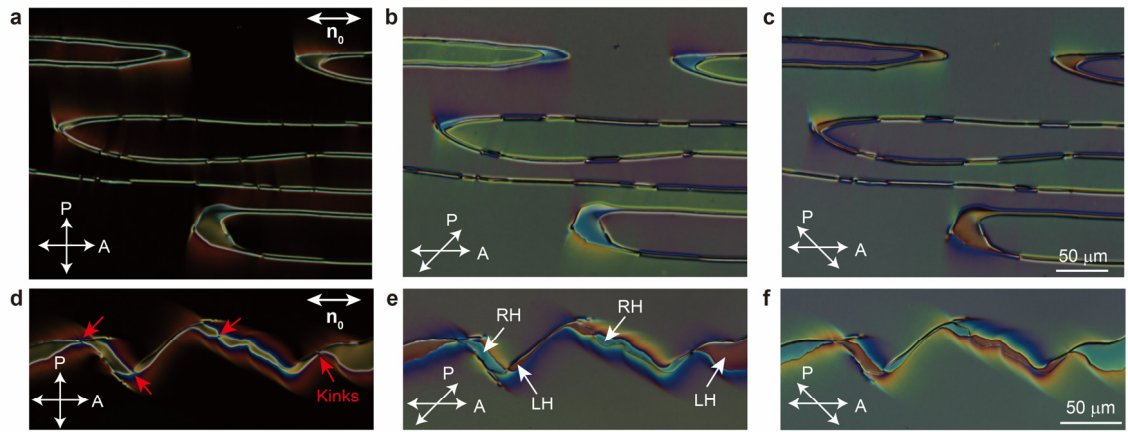

**Extended Data Fig.10|Separation of twin disclinations in photoaligned cells by shear.** (**a-c**) Polarized optical microscopy images when a small pressure was applied. (**d-f**) Polarized optical microscopy images when a relatively large pressure was applied. Subdomains exhibit left- (LH) and right-handed (RH) twists of polarizations across the cell. The cell thickness is 3 μm and filled with RM734.

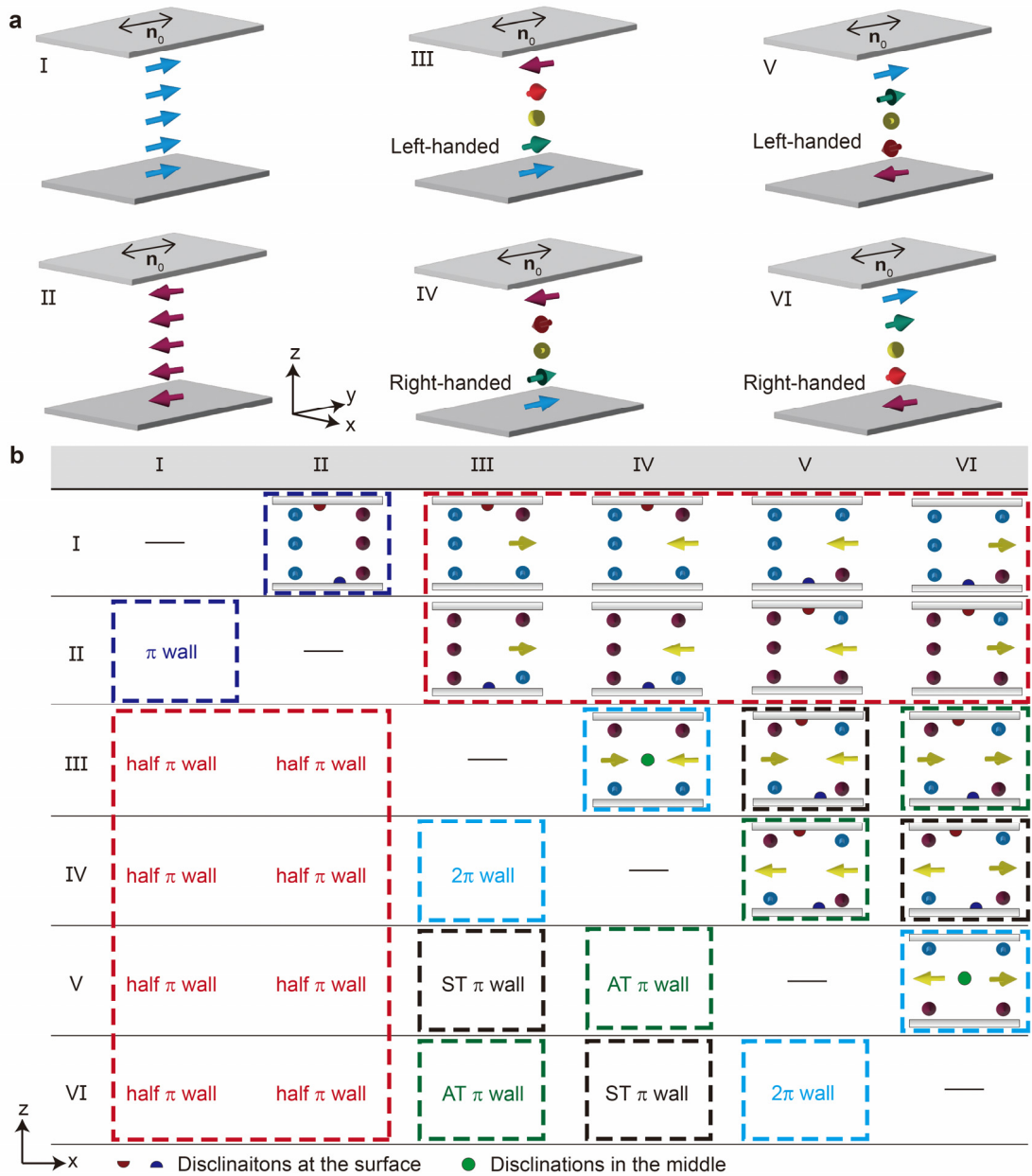

**Extended Data Fig.11|Possible domain walls in a uniform alignment cell.** (**a**) Six possible configurations that satisfy the boundary constraints. (**b**) Five types of domain walls separating two different domains in (**c**): π wall (dark-blue dashed box); half π wall (red); 2π wall (light blue), asynchronous-twist (AT) π wall (green), and synchronous-twist (ST) π wall (black).

**Video Captions**

**Video 1**

This video shows phase transitions of the $N_F$ liquid crystal DIO. The liquid crystal cell is made with rubbing method, and of $h = 4.4$ μm in thickness. The cell was cooled from the isotropic phase to the $N_F$ phase at a cooling rate of 5 $K$/min.

**Video 2**

This video shows the dynamics of $\pi$ walls under shear flow. The liquid crystal cell is made with the rubbing method, of $h = 5.1$ μm in thickness and filled with DIO. When a small pressure is applied to the cell, the induced flow causes the twin lines of $\pi$ walls to separate further. Liquid crystal molecules are strongly anchored to the substrates, so the flow-induced distortions of the polarization field give rise to restoring forces on twin lines. When the flow is off, this restoring torque drives the twin lines back into their original configuration. The video also demonstrates the creation and annihilation of kinks and antikinks.

**Video 3**

This video shows the dynamics of $\pi$ walls under shear flow. The liquid crystal cell is made by the photoalignment method, of $h = 3$ μm in thickness and filled with RM734. When a small pressure is applied, the shear flow causes further separations of the twin lines. When the flow is off, the recovering of original twin line spacings is relatively slow, compared to that in a rubbed cell.

**Video 4**

This video shows the polar switching driven by an in-plane electric field **E**. The cell is of $h = 3.3$ μm in thickness and filled with RM734. The field strength is gradually increased from 0 to 30 V/mm. The polar inversion undergoes two distinct steps. In the first step, two disclinations at a confining substrate are annihilated, leaving a twist domain. In the second step, the twist domain shrinks in size and disappears eventually.